\documentclass[english]{article}
\usepackage[T1]{fontenc}
\usepackage[latin9]{inputenc}
\usepackage{geometry}
\usepackage{color}
\usepackage{amsmath}
\usepackage{amssymb}
\usepackage{graphicx}
\usepackage[sort&compress,numbers]{natbib}
\usepackage{nohyperref}

\makeatletter
\usepackage{epsfig}
\usepackage{esint}
\usepackage{times,epstopdf}

\@ifundefined{showcaptionsetup}{}{%
 \PassOptionsToPackage{caption=false}{subfig}}
\usepackage{subfig}
\makeatother

\usepackage{babel}
\begin{document}
\title{  Continuous variable B92 quantum key distribution protocol
using single photon added and subtracted coherent states }
\author{  {\normalsize{}Srikara S$^{a}$, Kishore Thapliyal$^{b,}$}\thanks{Email: kishore.thapliyal@upol.cz}{\normalsize{},
Anirban Pathak$^{c,}$}\thanks{Email: anirban.pathak@jiit.ac.in}{\normalsize{}
}\\
{\normalsize{}$^{a}$Indian Institute of Science Education and Research,
Pune, India}\\
{\normalsize{}$^{b}$RCPTM, Joint Laboratory of Optics of Palacky
University and Institute of Physics of Academy of Science of the }\\
{\normalsize{}Czech Republic, Faculty of Science, Palacky University,
17. listopadu 12, 771 46 Olomouc, Czech Republic}\\
{\normalsize{}$^{c}$Jaypee Institute of Information Technology, A-10,
Sector-62, Noida, UP-201307, India}}
\maketitle
\begin{abstract}
  In this paper,  a continuous variable B92 quantum
key distribution (QKD) protocol is proposed using single photon added and subtracted
coherent states, which are prepared by adding and subsequently subtracting
a single photon on a coherent state. It is established that in contrast to the traditional
discrete variable B92 protocol, this protocol for QKD is intrinsically robust
against the unambiguous state discrimination attack, which circumvents
the requirement for any uninformative states or entanglement used
in corresponding discrete variable case as a remedy for this attack.
Further, it is shown that the proposed protocol  is intrinsically
robust against the eavesdropping strategies exploiting classical communication
during basis reconciliation, such as beam splitter attack. Security
against some individual attacks, key rate, and bit-error rate estimation
for the proposed scheme are also provided. Specifically, the proposed scheme ensures very small bit-error rate due to properties of the states used. Thus, the proposed scheme
is shown to be preferable over the corresponding discrete variable B92 protocol as well
as some similar continuous variable quantum key distribution schemes. 
\end{abstract}

\section{Introduction\label{sec:Introduction}}

  Quantum key distribution (QKD) is a method by which quantum
states and features of quantum mechanics are used to distribute an
unconditionally secure key (see \cite{gisin2002quantum,pathak2013elements,shenoy2017quantum}
for review). A scheme for QKD was first proposed by Bennett and Brassard
in 1984, which is now known as BB84 protocol for QKD \cite{bennett1984quantum}.
This was followed by an entangled state based protocol for QKD \cite{ekert1991quantum}
introduced by Ekert in 1991, which later formed the basis of device
independence. This protocol involved measurement in three bases and
thus uses six states in contrast to four states used in BB84 protocol.
Almost immediately after the introduction of Ekert's protocol, in
1992, Bennett established that neither four nor six states are essential
to accomplish the QKD task in a secure manner, and two non-orthogonal
states are sufficient to perform QKD \cite{bennett1992quantum}. The
protocol drew considerable attention of the community as it successfully
provided some fundamental insights into the origin of the 
security in the schemes of QKD. This scheme is now known as B92 protocol. In the original
form this protocol was designed for discrete variable (DV) QKD, here
we aim to extend it to continuous variable (CV) QKD for certain advantages
that CVQKD provides over its DV counterpart. Before we elaborate on the advantages and limitations
of CVQKD, it would be of use to briefly recall the original B92 protocol.

  In the B92 protocol \cite{bennett1992quantum}, Alice prepares
and sends a string of qubits individually prepared randomly in state
$|0\rangle$ (representing the bit value 0) or $|+\rangle=(|0\rangle+|1\rangle)/\sqrt{2}$
(representing the bit value 1) to Bob. Bob measures each incoming
qubit randomly either in the computational $\left\{ |0\rangle,|1\rangle\right\} $
or diagonal $\left\{ |+\rangle,|-\rangle\right\} $ basis, and designates
the measured bit as 0 (1) if his measurement outcome is $|-\rangle$
($|1\rangle$). For the measurement outcomes $|0\rangle$ or $|+\rangle$,
he declares the result as inconclusive, and discards these results.
A part of the remaining conclusive bits are used to check eavesdropping
and the rest are retained by both Alice and Bob, thus sharing a symmetric
secret key. This protocol is unconditionally secure under an ideal
lossless channel. However, it can be attacked by Eve in a lossy channel
using the unambiguous state discrimination (USD) attack \cite{chefles1998unambiguous,dusek2006quantum},
which makes this protocol insecure in extremely lossy channels, thus
making it difficult to be practically implemented. A modification
of the B92 protocol, which was robust against the USD attack was introduced
in \cite{lucamarini2009robust}, using a non-maximally entangled Bell
pair and two additional decoy states, called the uninformative states.
A device independent version of the B92 protocol \cite{lucamarini2009robust}
was also proposed in the recent past \cite{lucamarini2012device}
by allowing the entanglement generation in the hands of an untrusted
party present between Alice and Bob. Recently, B92 protocol was found
relevant in designing a quantum private query scheme \cite{yang2014flexible},
which was later shown insecure in lossy channel \cite{chang2017comment}.
Further, an advanced USD attack is also designed for cryptanalysis
of B92 scheme \cite{ko2018advanced}.

In contrast to the above mentioned DV quantum cryptography schemes,
where photon counters are used, a set of CV quantum communication
schemes is also proposed in which information is encoded on quadratures
and decoded by homodyne or heterodyne detection (see \cite{braunstein2005quantum,andersen2010continuous,weedbrook2012gaussian}
for review). The first set of CV cryptography schemes was proposed
using squeezed state \cite{hillery2000quantum}, Einstein-Podolsky-Rosen
correlated states \cite{reid2000quantum}, and coherent state \cite{ralph1999continuous,ralph2000security,PhysRevA.63.022309,grosshans2002continuous}.
The initial set of proposals was focused on encoding discrete
quantum key in CV quantum states, such hybrid schemes were  later
extended to all CV schemes distributing CV quantum key \cite{cerf2001quantum}.
The CV quantum cryptography has several advantages over its DV counterpart
as it involves multi-photon pump beams and do not require single photon
detectors, which omit the limitations of single photon source and
detector. On top of that, implementation with easily available light
source(s) and detector(s), compatibility with the existing optical communication
technology, and advantage of CV schemes at short distances make them
preferred candidates for metropolitan secure networks. Historical
development and  current status of the CV cryptography can be found
in a set of review articles \cite{braunstein2005quantum,lutkenhaus2009focus,weedbrook2012gaussian,diamanti2015distributing}.
Because of these advantages of CVQKD, a CV counterpart of the BB84 scheme
\cite{bennett1984quantum} has recently been proposed using 
single photon added and subtracted coherent states (PASCS) \cite{borelli2016quantum}.
This scheme uses four states analogous to the corresponding DV BB84 scheme
and also involves classical communication at the end to discard the
cases where their choices of quadratures are different (see Ref. \cite{borelli2016quantum}
for more detail). The security of CV QKD schemes is analyzed in detail
(see \cite{scarani2009security,lutkenhaus2009focus} for review).
For instance, different studies have reported security against general attack \cite{leverrier2017security},
composable security \cite{leverrier2015composable}, machine learning
for parameter estimation \cite{liu2018integrating}, using whole raw
key for parameter estimation without compromising security \cite{lupo2018parameter}.
On the other hand, initial CV QKD schemes have also led to the schemes for measurement device independent
\cite{pirandola2015high}, device independent \cite{marshall2014device},
entanglement-distribution-based \cite{zhou2018long}, atmospheric
\cite{heim2014atmospheric}, satellite-based \cite{hosseinidehaj2018satellite}
CV-QKD as well as hybrid CV- and DV-QKD \cite{liu2015hybrid}. More
recently, proposal to completely eliminate information leakage in
CV communication performed over lossy channels has been reported \cite{jacobsen2018complete},
which has no DV analogue. The proposals for CV cryptography are not
restricted to QKD as CV schemes for quantum signature \cite{croal2016free},
direct secure quantum communication  \cite{saxena2019continuous},
quantum secret sharing \cite{wu2016continuous}, and position-based
quantum cryptography \cite{qi2015loss} are also proposed. 

  Motivated by the above facts, in this paper, we propose
a CV version of the B92 protocol \cite{bennett1992quantum},
which can be viewed as a B92 type modification of a QKD protocol \cite{borelli2016quantum}.
The protocol proposed in this paper is intrinsically robust against
the USD attack, i.e., it does not require any uninformative state
or entanglement. Also, unlike the protocol discussed in Ref. \cite{borelli2016quantum},
the proposed protocol is intrinsically (i.e., without any additional
conditions on any parameters) robust against the beam splitter attack,
too. Such discrete modulation CV-QKD schemes are shown to perform
better by using efficient error correction codes \cite{ma2019long}.

  The rest of the paper is structured as follows. In Section
\ref{sec:Photon-Added-and}, we briefly introduce PASCS and the mathematical
tools that are used in this work. Section \ref{sec:CV-B92-Protocol}
describes the protocol, and Section \ref{sec:Security-Analysis-of}
is dedicated to the security analysis of the proposed protocol. Finally, the paper is concluded
in Section \ref{sec:Conclusion}. 

\section{Photon added and subtracted coherent states\label{sec:Photon-Added-and}}

  In this section, we aim to introduce the mathematical tools
and the quantum states that will be required to explain the protocol.
To begin with we may note that coherent state, which is usually obtained
as the state of the radiation field at the output of a laser source,
is essentially an eigen-ket of the annihilation operator. It can also
be described as a displaced vacuum state. Subsequent photon addition
and subtraction on a coherent state leads to PASCS. This state seems
physically realizable as the photon addition and subtraction operations
are experimentally feasible (see \cite{parigi2007probing,thapliyal2017comparison,malpani2018lower}
and references therein). A single photon added then subtracted coherent
state $|\psi\left(\gamma\right)\rangle$ is defined as

\begin{equation}
|\psi\left(\gamma\right)\rangle=N_{\gamma}^{-1/2}\hat{a}\hat{a}^{\dagger}|\gamma\rangle,\label{eq:pascs}
\end{equation}
where $|\gamma\rangle$ is the initial coherent state having an average
photon number $\left|\gamma\right|^{2}$ with $a$ and $a^{\dagger}$
corresponding to annihilation and creation operators, respectively.
Further, $N_{\gamma}$ is the normalization constant which can be
expressed as

\[
N_{\gamma}=|\gamma|^{4}+3|\gamma|^{2}+1.
\]
In what follows, we propose a protocol of QKD using two PASCSs described by Eq, (\ref{eq:pascs}) and 
characterized by $\gamma=\alpha$ (i.e., $|\psi_{0}\left(\alpha\right)\rangle$)
and $\gamma=i\alpha$ (i.e., $|\psi_{1}\left(\alpha\right)\rangle$)
for $\alpha\in\mathbb{R}^{+}$. 

  A phase space description of quantum mechanics was introduced
by Wigner in 1932 \cite{wigner1932quantum}. As this distribution
function can have negative values (normalized to unity), it is not
a true probability distribution function, and is often referred to
as quasidistribution function. Much later, Glauber and Sudarshan introduced
the notion of nonclassicality in terms of negative values of $P$
function \cite{glauber1963coherent,sudarshan1963equivalence}, and
it was realized that the states having negative values of Wigner function
must be nonclassical. The Wigner function for an arbitrary state,
with density matrix $\rho$, is given by \cite{moya1993series}

\begin{equation}
W(\zeta)=\frac{2}{\pi}\overset{\infty}{\underset{n=0}{\sum}}(-1)^{n}\langle n|\hat{D}^{-1}(\zeta)\hat{\rho}\hat{D}(\zeta)|n\rangle,\label{eq:wigner}
\end{equation}
where $\hat{D}(\zeta)=e^{(\zeta\hat{a}^{\dagger}-\zeta^{*}\hat{a})}$
is the displacement operator defined in terms of complex number $\zeta=\zeta_{r}+i\zeta_{i}$.
Here $\zeta_{r}$ and $\zeta_{i}$ are the phase space coordinates
corresponding to the position and momentum quadratures, respectively,
and $|n\rangle$ is the Fock state. Now, using Eq. (\ref{eq:wigner}),
we can obtain Wigner function for the PASCS of our interest. Specifically,
Wigner function for $|\psi_{j}\left(\alpha\right)\rangle$ is obtained
as

\begin{equation}
W_{j}\left(\zeta,\alpha\right)=\frac{2e^{-2\left[\left(\zeta_{a}-\alpha\right)^{2}+\zeta_{b}^{2}\right]}}{\pi\left(\alpha^{4}+3\alpha^{2}+1\right)}\left\{ \left(\alpha^{2}-1-2\zeta_{a}\alpha\right)^{2}+\left(2\zeta_{b}\alpha\right)^{2}-\alpha^{2}\right\} ,\label{eq:Wig-PASCS0}
\end{equation}
where $\left\{ a=r,b=i\right\} $ for $j=0$, while $\left\{ a=i,b=r\right\} $
for $j=1$. 

  It is already mentioned in the previous section that a CV
quantum communication scheme is different from a DV scheme designed
for the same task as it involves homodyne measurement(s). This can be
performed by mixing the quantum signal with a classical beam at a
beam splitter and measuring the difference of the currents at two output
ports. By controlling the phase of the input classical beam, we can
address one of the quadratures $\zeta_{r}$ (i.e., position) or $\zeta_{i}$
(i.e., momentum).

\begin{figure}
\centering{}\subfloat[]{\begin{centering}
\includegraphics[scale=0.6]{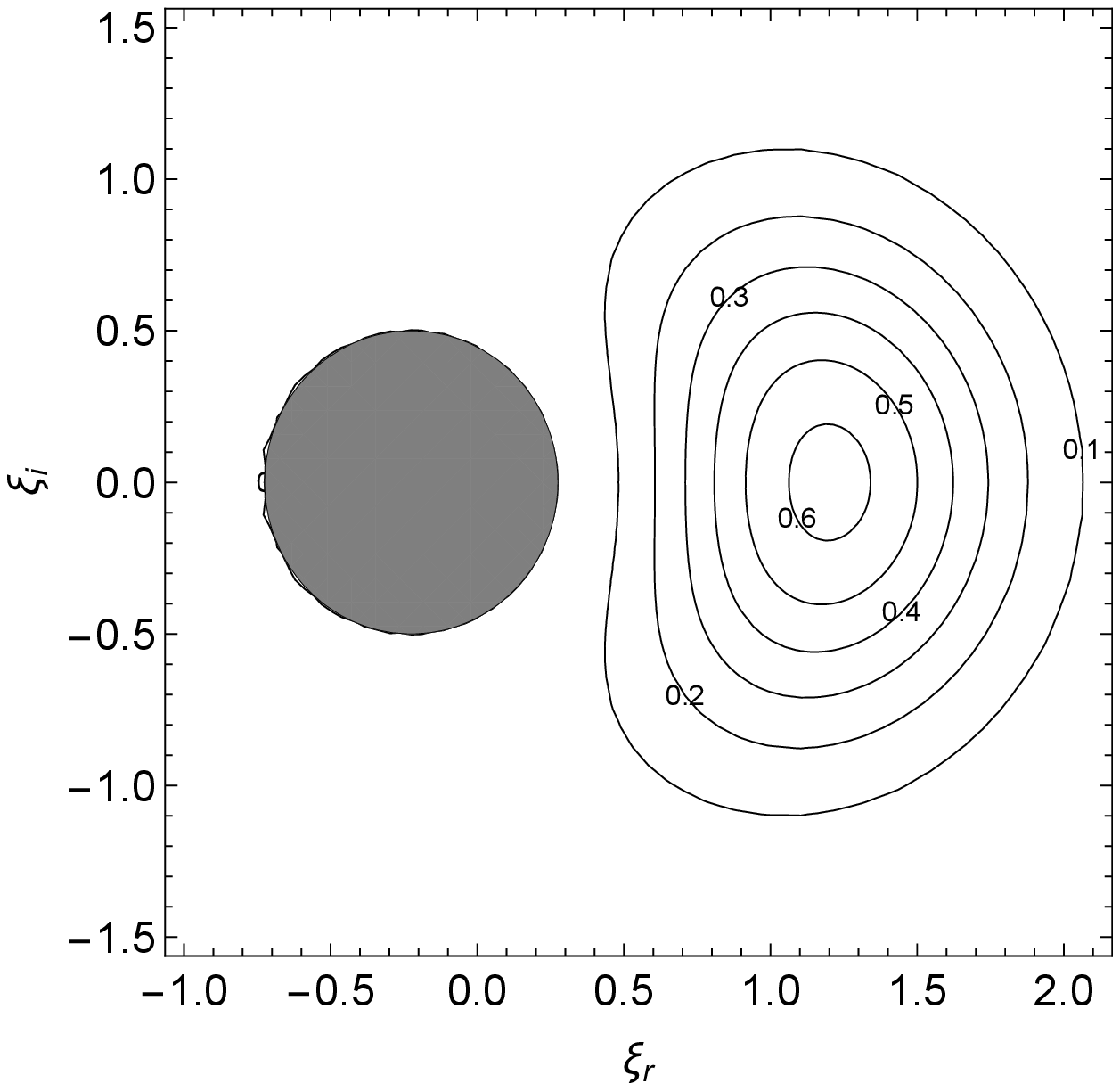} 
\par\end{centering}

}\subfloat[]{\begin{centering}
\includegraphics[scale=0.6]{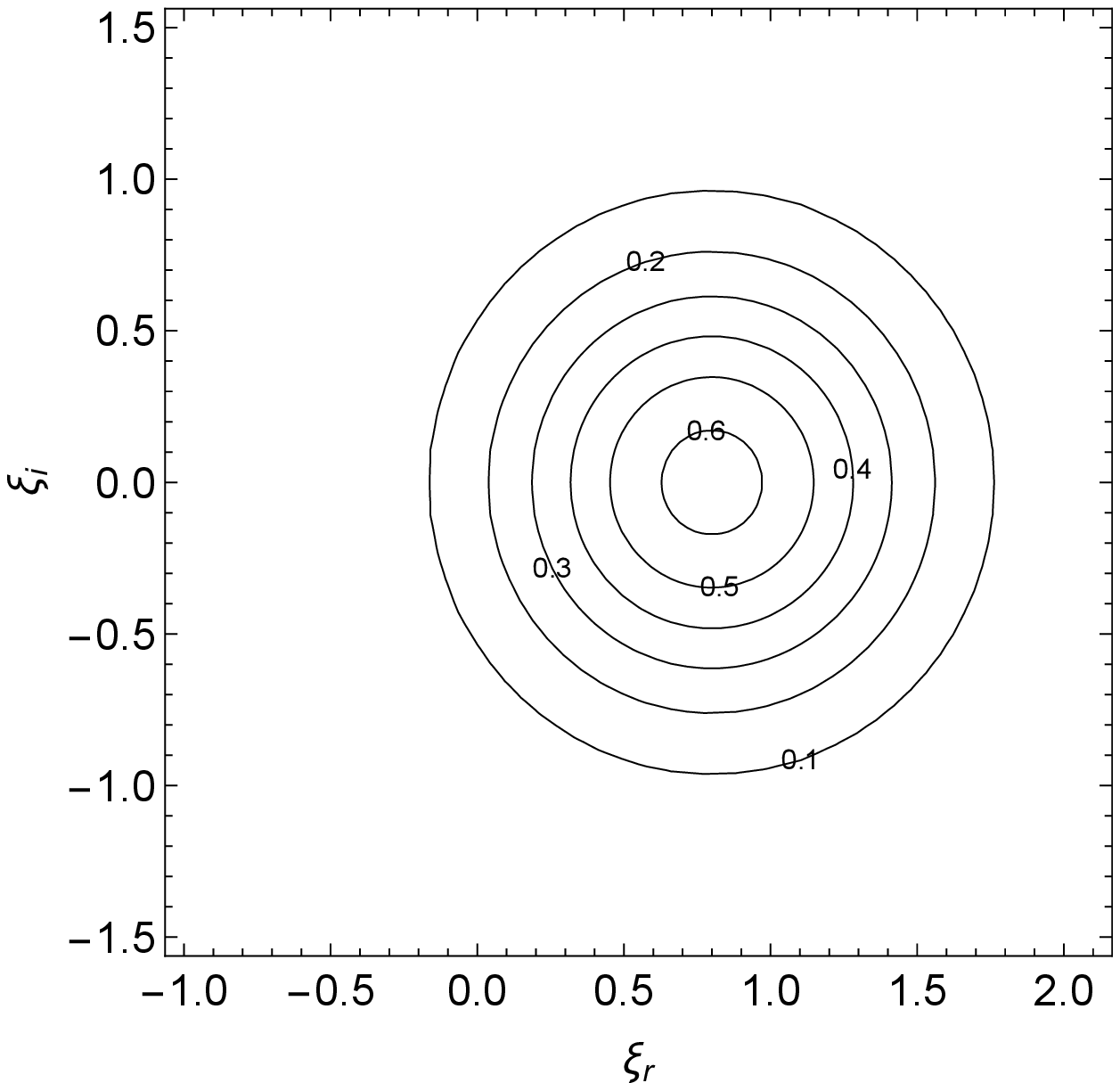}
\par\end{centering}

}\caption{\label{fig:Wigner-functions} Contour plots of Wigner functions of (a)
PASCS $|\psi_{0}\left(\alpha\right)\rangle$ and (b) the corresponding
coherent state with $\alpha=0.8$. The gray region in (a) shows the
negative values of Wigner function and thus provides a signature of nonclassicality. }
\end{figure}
  For the sake of completeness, we have also shown the surface
plots of the Wigner functions of the PASCS $|\psi_{0}\left(\alpha\right)\rangle$
and coherent state $|\alpha\rangle$ with $\alpha=0.8$ in Fig. \ref{fig:Wigner-functions}
(a) and (b), respectively. Clearly, the Wigner function of the PASCS
has a negative region (shown by gray color in Fig. \ref{fig:Wigner-functions}
(a)), indicating that the state is nonclassical. Due to applications
of two non-Gaussianity inducing operations, namely photon addition
and subtraction, in generation of PASCS, the obtained Wigner function
is non-Gaussian and also shows a shift toward the positive real side
as compared to the corresponding coherent state. In a homodyne detection,
one may choose either to measure quadrature $\zeta_{r}$ or quadrature
$\zeta_{i}$. In analogy of Fig. \ref{fig:Wigner-functions}
(a) obtained for $|\psi_{0}\left(\alpha\right)\rangle$, one can also obtain the Wigner function for PASCS
$|\psi_{1}\left(\alpha\right)\rangle$. In what follows, we will propose
our CV B92 QKD scheme using PASCS described in this section. 

\section{Continuous Variable B92 protocol\label{sec:CV-B92-Protocol}}

  Our protocol 
described below is a CV counterpart of B92 protocol. Alice and Bob
initially publicly agree upon a positive real number $\alpha$. Depending
upon that Bob fixes another positive real value for the post selection
threshold $\zeta_{c}$ in the beginning. The protocol can be described
as follows.
\begin{enumerate}
\item Alice prepares an $n$ bit random string ($K_{A}$) and prepares signal
pulses in PASCS either as $|\psi_{0}\left(\alpha\right)\rangle$ corresponding
to the bit value 0 or $|\psi_{1}\left(\alpha\right)\rangle$ for bit
value 1 in $K_{A}$. Subsequently, she sends the sequence $S_{A}$
of the prepared states to Bob.
\item Bob also prepares a random string $K_{B}$ of $n$ bit. Corresponding
to the bit value 0 (1) in $K_{B}$, he selects to measure the position
(momentum) quadrature of the signal pulses in $S_{A}$ in the homodyne
measurement resulting in the real (imaginary) part of $\zeta$.
\item If Bob performs homodyne measurement on the position (momentum) quadrature
and obtains $\zeta_{r}\,<\,-\zeta_{c}$ ($\zeta_{i}\,<-\zeta_{c}$)
he designates it as conclusive. Otherwise, he discards the result
and terms the corresponding outcome as inconclusive.
\item Bob then declares the coordinates of the retained (conclusive) results,
using which both Alice and Bob obtain $K_{A}^{R}\in K_{A}$ and $K_{B}^{R}\in K_{B}$,
respectively, after discarding the bit values corresponding to inconclusive
results. Among these conclusive results, Bob takes a part of it and
announces his measurement outcomes and the coordinates of this part.
Alice and Bob then perform the eavesdropping checking on these results,
i.e., they count the number of instances when Alice sent 0 (1) and Bob
obtained 1 (0). If the error is within a tolerable limit, they continue
to Step 5, else they discard the protocol and start afresh.
\item Alice and Bob discard the results used for eavesdropping checking
and retain the remaining conclusive results $K_{A}^{f}\in K_{A}^{R}$
and $K_{B}^{f}\in K_{B}^{R}$, respectively, hence obtaining a shared
secret key. 
\end{enumerate}
At the end of the quantum communication Alice and Bob are expected
to share an unconditionally secure quantum key, but in ideal conditions
$K_{B}^{f}=\overline{K}_{A}^{f}$. Therefore, it is predecided that
at the end of Step 5, Bob will flip his key to ensure $\overline{K}_{B}^{f}=K_{A}^{f}=K$. 

Bit error rate (when $K_{B}^{f}=K_{A}^{f}$) is expected to be low 
due to negligibly small non-zero value of marginal distribution in the region $\zeta_{r}\,<\,-\zeta_{c}$
for $|\psi_{0}\left(\alpha\right)\rangle$.  However, this can be circumvented
by using error correction and privacy amplification \cite{ma2019long,lutkenhaus2009focus}, we will discuss this in the next section. 

\section{Information gain per transmitted state}

In this section, we are going to calculate the average amount of information
$G_{ab}$ (in bits) gained transferred Alice and Bob every time Alice
sends a PASCS through a lossy channel \cite{horak2004role}, which can
be modeled as passing through a beam splitter with transitivity $T$
and reflectivity $R$. Without any loss of generality, all the losses
can be attributed to the eavesdropping attempts by Eve. 

Suppose Alice transmits $|\psi_{j}(\alpha)\rangle$ to Bob. The Wigner
function of the attenuated signal in this model can be described
by two mode Wigner function of the state after $|\psi_{j}(\alpha)\rangle$
passes through the beam splitter as  
\begin{equation}
\widetilde{W}_{j}(\zeta,\epsilon)=W_{j}(T\zeta-R\epsilon,\alpha)\times W_{j}(R\zeta+T\epsilon,0).\label{eq:lossy}
\end{equation}
  Here, we assume the other input of the beam splitter as a single
mode vacuum state. Also, $W_{j}(\zeta,\alpha)$ is the Wigner function
for the PASCS states sent by Alice corresponding to $j=0$ and 1 defined
in Eq. (\ref{eq:Wig-PASCS0}). Using this the joint probability distribution
can be computed as \cite{borelli2016quantum,horak2004role}

\begin{equation}
P_{j}(\zeta_{x})=\int\widetilde{W}_{j}(\zeta,\epsilon)d\zeta_{y}d^{2}\epsilon,\label{eq:int-re}
\end{equation}
where $x,y\in\left\{ r,i\right\} :x\neq y$. It can further be used
to calculate the probability $P_{j}$ that Bob correctly infers the
state $|\psi_{j}(\alpha)\rangle$ sent by Alice (i.e., gets bit $j$) 

\begin{equation}
P_{j}=\overset{-\zeta_{c}}{\underset{-\infty}{\int}}P_{j}(\zeta_{b})d\zeta_{b},\label{eq:PR}
\end{equation}
while the probability $P_{\bar{j}}$ that Bob wrongly infers the state
(i.e., gets bit $\bar{j}$) is

\begin{equation}
P_{\bar{j}}=\overset{-\zeta_{c}}{\underset{-\infty}{\int}}P_{j}(\zeta_{a})d\zeta_{a},\label{eq:PW}
\end{equation}
where notation is same as in Eq. (\ref{eq:Wig-PASCS0}). Hence, the
fraction of accepted bits $r_{{\rm acc}}$ is given by

\begin{equation}
r_{{\rm acc}}=\frac{P_{j}+P_{\bar{j}}}{2}\label{eq:racc}
\end{equation}
with factor $1/2$ corresponding to the probability that $i$th term
in $K_{A}$ and the same in $K_{B}$ are the same. Further, the bit-error rate $\delta$
per conclusive result can be given as

\begin{equation}
\delta=\frac{P_{\bar{j}}}{P_{j}+P_{\bar{j}}}.\label{eq:BER}
\end{equation}

\begin{figure}
\centering{}
\subfloat[]{
\includegraphics[scale=0.45]{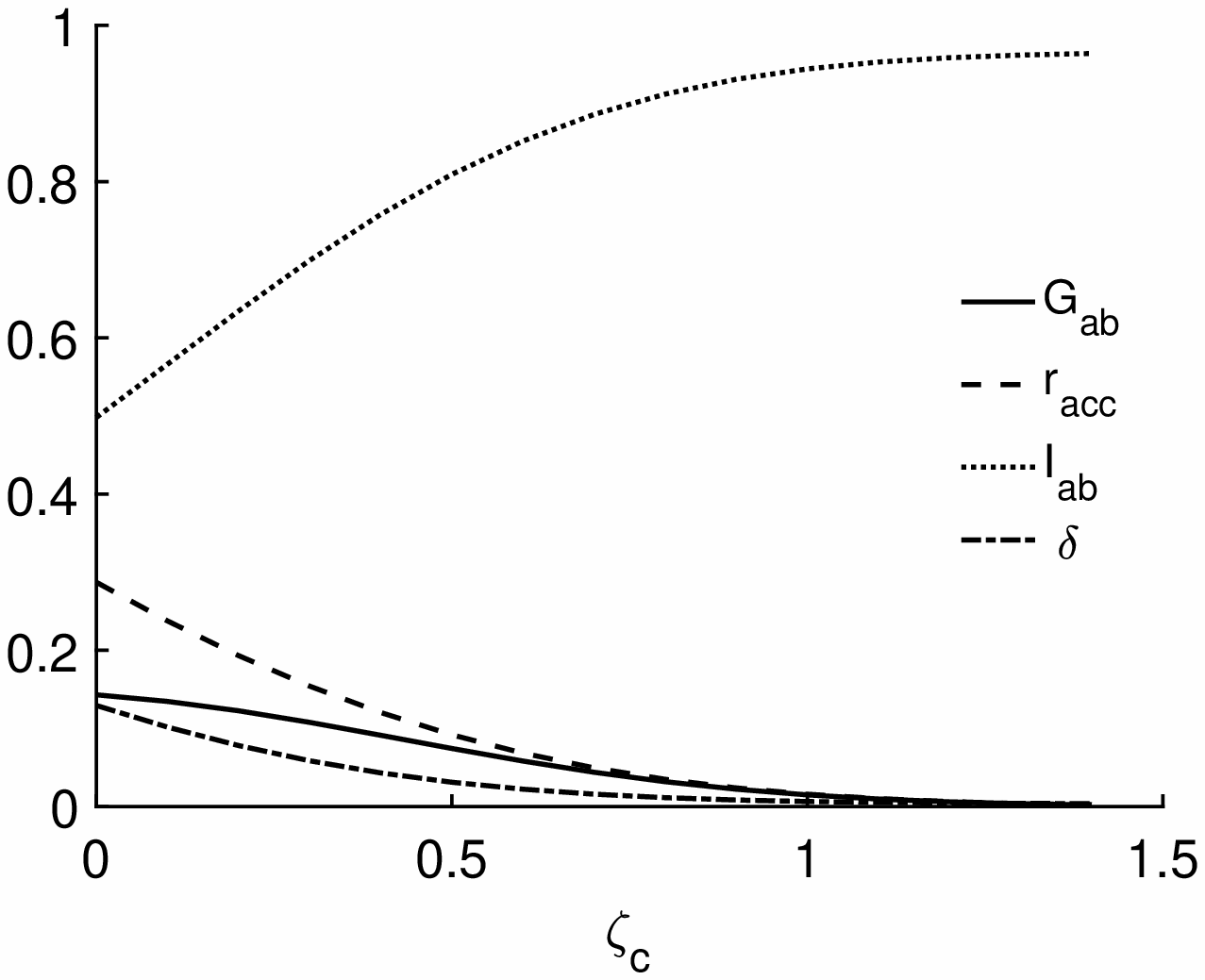} }
\subfloat[]{
\includegraphics[scale=0.45]{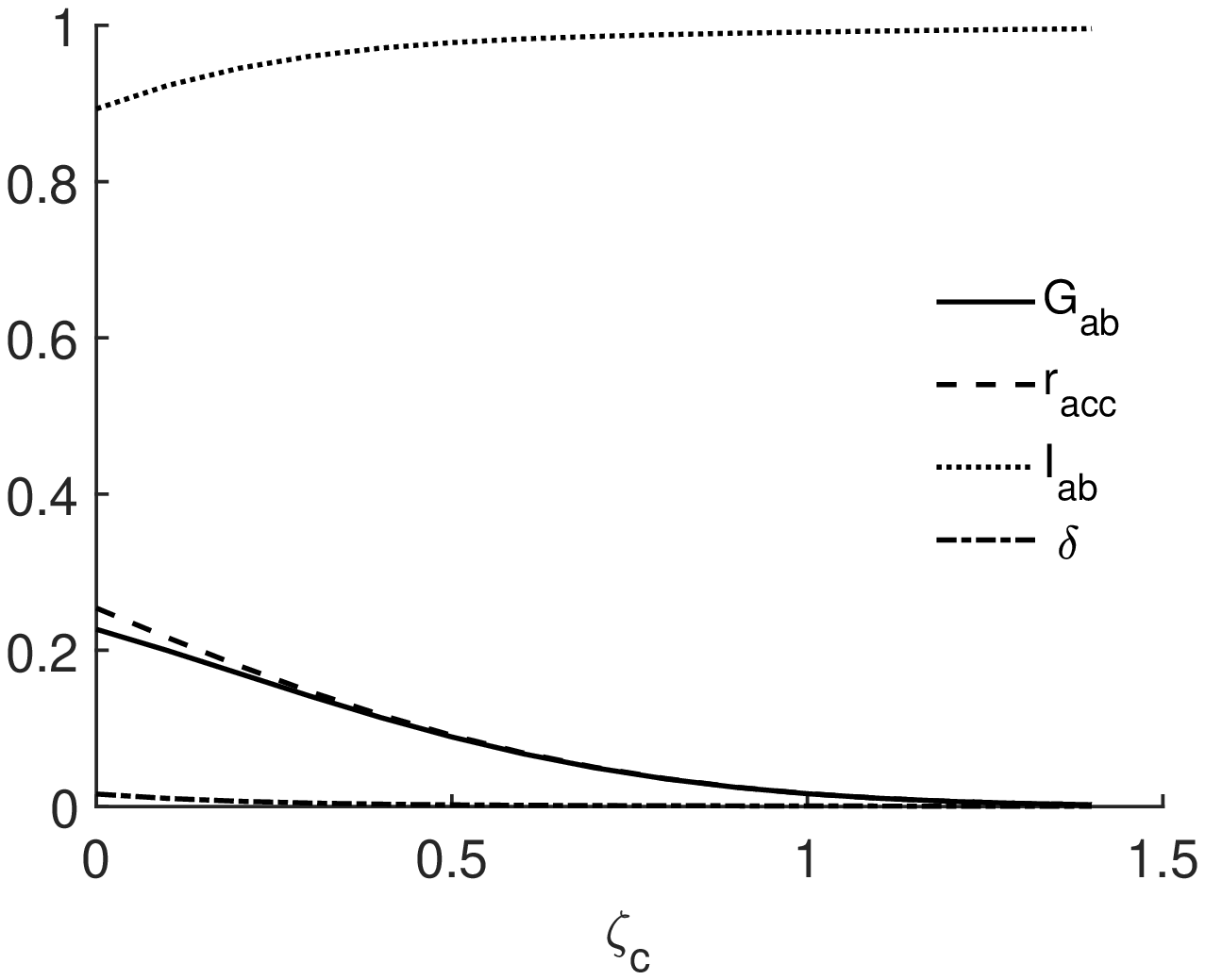} }
\subfloat[]{
\includegraphics[scale=0.45]{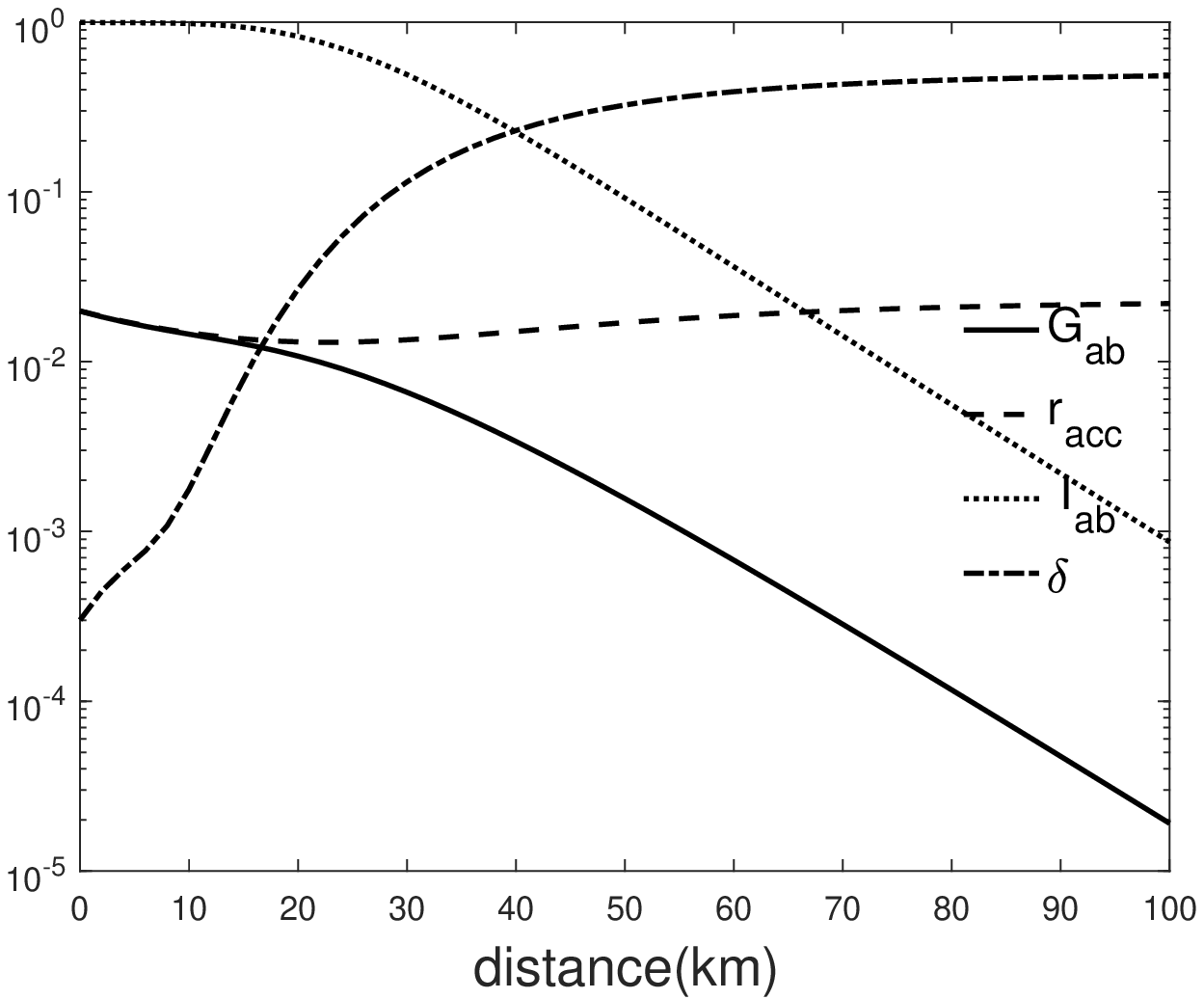}
}
 \caption{\label{fig:KeyRate} (a) Variation of fraction of accepted
bits $r_{{\rm acc}}$, mutual information $I_{ab}$, average information
$G_{ab}$, and bit-error rate $\delta$ with $\zeta_{c}$
considering $T=0.7$ and  $\alpha=0.6$.  All the parameters for $\alpha=1.2$ are shown as function of (b) threshold $\zeta_{c}$
considering $T=0.7$ and (b) distance considering $\zeta_{c}=1$.}
\end{figure}

Therefore, Shannon information between Alice and Bob per conclusive
result can be calculated by averaging over Bob's measurement outcomes

\begin{equation}
I_{ab_{j}}=\overset{-\zeta_{c}}{\underset{-\infty}{\int}}d^{2}\zeta\frac{P_{j}(\zeta_{b})+P_{j}(\zeta_{a})}{P_{j}+P_{\bar{j}}}\left\{ 1+\Phi(\zeta)\log_{2}\Phi(\zeta)+(1-\Phi(\zeta))\log_{2}(1-\Phi(\zeta))\right\} ,\label{eq:mut-inf}
\end{equation}
where error function $\Phi(\zeta)=\frac{P_{j}(\zeta_{a})}{P_{j}(\zeta_{b})+P_{j}(\zeta_{a})}$.
Thus, the average information $G_{ab}$ gained by Bob per transmitted
state by Alice is

\begin{equation}
G_{ab}=I_{ab}r_{{\rm acc}}.\label{eq:gain}
\end{equation}
Privacy amplification of the shared key provides a lower bound to
the secret information transmitted in one pulse as \cite{lutkenhaus1996security}
\begin{equation}
S_{ab}=r_{{\rm acc}}\left(I_{ab}-\tau\right)\label{eq:priv}
\end{equation}
at the cost of reduction of the key size by $\tau=1+\log_{2}P_{{\rm coll}}$,
where collision probability is
\begin{equation}
P_{{\rm coll}_{j}}=\int d^{2}\epsilon\frac{P_{j}(\epsilon|\zeta_{c}<\left|\zeta_{b}\right|)^{2}+P_{\bar{j}}(\epsilon|\zeta_{c}<\left|\zeta_{a}\right|)^{2}}{P_{j}(\epsilon|\zeta_{c}<\left|\zeta_{b}\right|)+P_{\bar{j}}(\epsilon|\zeta_{c}<\left|\zeta_{a}\right|)}\label{eq:coll-pr}
\end{equation}
with $P_{J}(\epsilon|\zeta_{c}<\left|\zeta_{x}\right|)=\overset{-\zeta_{c}}{\underset{-\infty}{\int}}\frac{P_{J}(\zeta_{x},\epsilon)}{P_{j}+P_{\bar{j}}}d\zeta_{x}$.
Variation of fraction of accepted bits $r_{{\rm acc}}$, mutual information
$I_{ab}$, average information $G_{ab}$, and bit-error rate $\delta$
with threshold value $\zeta_{c}$ are shown in Fig. \ref{fig:KeyRate} (a) and (b) which shows
similar variation as reported in Ref. \cite{horak2004role}. Here, the noteworthy point is extremely
low bit-error rate provided by the present scheme (cf. Fig. \ref{fig:KeyRate} (b)). Further, smaller value of
$r_{{\rm acc}}$, and thus $G_{ab}$, can be attributed to the fact
that due to very small errors, reflected through small bit-error rate, contribution of wrong accepted bits is low as well as only non-unity value of transmitivity $T$ is considered here. However, the larger value
of information transmitted per accepted bit is due to the high value
of $\alpha$ (to visualize this compare Figs. \ref{fig:KeyRate}
(a) and  (b).  The same set of parameters is also shown as functions
of the transmission distance over an optical fiber with absorption
rate 0.02 and Bob's homodyne detection efficiency 0.9 in Fig. \ref{fig:KeyRate}
(c), where one can clearly see the increase in the bit-error rate and
decrease in both information transmitted per accepted bit and average
information shared with losses in the channel. Finally, security of the extracted key can be enhanced by reducing the size of key in privacy amplification.

\section{Security analysis of the protocol\label{sec:Security-Analysis-of}}

  The security of the proposed QKD protocol against an adversary
will be discussed under three specific attacks. We begin with the
beam splitter attack in which Eve exploits the transmission losses
by using a beam splitter. In another attack, Eve may intercept the
transmitted signal pulses to measure both the quadratures. Finally,
we will consider a particular attack referred to as USD attack which
was successfully used to attack the original B92 protocol (DV version
of our scheme). 

\subsection{Beam splitter attack}

  In the context of CVQKD, a commonly discussed attack strategy
is the beam splitter attack or superior channel attack \cite{horak2004role,namiki2004practical,borelli2016quantum}.
In the this attack, Eve uses a beam splitter to split the incoming
signal pulses from Alice to Bob. She sends one of the outputs of the
beam splitter to Bob, while keeps the other output in a quantum memory.
Subsequently, during quadrature reconciliation at the end of quantum
transmission, Bob announces his choice of quadrature measurement and
keeps only the outcomes when both Alice and Bob had chosen same quadratures.
During this step, Eve also comes to know the choice of quadrature
by Alice and Bob. Subsequently, Eve performs measurement of the same
quadratures in the corresponding pulses to infer the bit values shared
by Alice and Bob. 

  It is easy to show that the protocol described in Section
\ref{sec:CV-B92-Protocol} is robust against the above mentioned attack.
This is so because, to succeed in the beam splitter attack, Eve needs
access to the classical information that Alice and Bob share during
quadrature reconciliation. In the proposed protocol, there is no need
for this classical communication at the end of the measurement, so
Eve cannot obtain the information accessible to her in the CV-BB84
scheme \cite{borelli2016quantum} and looses the advantage of the
beam splitter attack. This observation also establishes supremacy of the proposed CV B92 protocol over the earlier proposed CV-BB84 protocol \cite{borelli2016quantum}.

\subsection{Intercept and resend attack}

Though beam splitter attack is not useful for Eve in the present case,
she can adopt different strategies. For example, she may attempt an
intercept and resend attack or simultaneous quadrature measurement
attack \cite{borelli2016quantum,horak2004role}. In this attack, Eve
uses a symmetric beam splitter (with $R=T$) to perform homodyne measurements
of two different quadratures (say $\beta_{r}$ and $\epsilon_{i}$)
in the two outputs of the beam splitter. Using the measurement outcomes,
she prepares and sends a state to Bob, for which the joint probability
distribution $P_{j}(\beta_{r},\epsilon_{i})$ is maximum. Both $|\psi_{0}\left(\alpha\right)\rangle$
and $|\psi_{1}\left(\alpha\right)\rangle$ have independent regions
in phase space with $\beta_{r}\,\ge\,|\epsilon_{i}|$ and $\epsilon_{i}\,>\,|\beta_{r}|$
of area $A_{0}$, respectively. The probability that Eve will infer
the state successfully, given the incoming state, $P_{\mathrm{corr}}$
can be computed using Eq. (\ref{eq:int-re}) as

\begin{equation}
P_{\mathrm{corr}}=\int_{A_{0}}P_{j}(\beta_{r},\epsilon_{i})d\beta_{r}d\epsilon_{i}.\label{eq:suc}
\end{equation}

Initially, before starting the protocol, Bob chooses a post-selection threshold
$\zeta_{c}$ and Alice chooses the optimum $\alpha$ for a given fixed
bit-error rate $\delta$. Eve sends the correct PASCS only with a
probability $P_{\mathrm{corr}}$ to Bob. Hence, lesser the value of
$P_{\mathrm{corr}}$, higher the probability of Eve getting detected
by a significant change in the bit-error rate $\delta$ or the rate
$r_{{\rm acc}}$ of accepted bits or by a change in the probability
distribution of Bob's quadrature measurements.

\subsection{Unambiguous state discrimination (USD) attack}

  This is a version of the intercept and resend attack which
is known to be a major drawback for the DV B92 protocol
\cite{dusek2006quantum}. In the DV B92 protocol, Eve
intercepts all the transmitted qubits and measures the state sent
by Alice randomly either in the computational $\left(Z\right)$ basis
or diagonal $\left(X\right)$ basis. If she obtains a conclusive result,
she re-prepares and sends the freshly prepared state to Bob by an
ideal channel. On the other hand, if she obtains an inconclusive result,
she does not resend anything and thus mimics a lossy channel. This
attack cannot be performed in the proposed CV scenario because in
the CV case, Bob performs homodyne measurement instead of the polarization
measurement or the photon number measurement. Hence, he will always
expect some signal coming towards him instead of vacuum. If Eve performs
the USD attack on the protocol described in Section \ref{sec:CV-B92-Protocol}
and finds an inconclusive result, she would  send a vacuum state and will
easily get caught as Bob will not receive any signal. Therefore, this
protocol is robust against the USD attack without any requirement
of using an alternate basis measurements, uninformative states or
entanglement \cite{lucamarini2009robust,lucamarini2012device}.

\section{Conclusion\label{sec:Conclusion}}

  A CV counterpart of B92 protocol is proposed here using
PASCS. The proposed protocol is shown to be free from the limitations of the DV B92 protocol, which is  prone
to USD attack. Due to this fact that the proposed B92 scheme omits
the requirement for any uninformative states or entanglement. On top
of that, the proposed protocol is also resistant to all the eavesdropping
strategies which exploit classical communication during basis reconciliation;
for instance, it is intrinsically robust against the beam splitter
attack. Therefore, the proposed scheme provides security against some
of the individual attacks and omits requirement of a classical communication
channel without compromising with the key rate and bit-error rate
estimated for other similar CV QKD schemes \cite{borelli2016quantum,horak2004role}.
We have also established the security of our scheme against other
individual attacks such as intercept and resend attack. 

  Security of our scheme over collective attacks remains an
open problem and will be attempted in the future. The proposed scheme
helps in sharing a discrete key between two parties by encoding it
on a continuous quantum carrier and can also be extended to design an all continuous
B92 scheme where continuous key can be shared with the help of a continuous
quantum carrier \cite{cerf2001quantum}.  With the well-known advantages of CVQKD schemes over DVQKD schemes, this protocols adds some additional benefits (robustness against certain attacks), and thus provides a potential scheme for practical implementation. 

  \textbf{Acknowledgments} AP acknowledges the support from Interdisciplinary Cyber   Physical   Systems   (ICPS)   programme   of   the Department  of  Science  and  Technology (DST),  India,  Grant No.:DST/ICPS/QuST/Theme-1/2019/14. KT
acknowledges the financial support from the Operational Programme Research, Development and Education - European Regional Development Fund project no. CZ.02.1.01/0.0/0.0/16 019/0000754 of the Ministry of Education, Youth and Sports of the Czech Republic.

\bibliographystyle{apsrev4-1}
\bibliography{refs}

\begin{thebibliography}{53}%
\makeatletter
\providecommand \@ifxundefined [1]{%
 \@ifx{#1\undefined}
}%
\providecommand \@ifnum [1]{%
 \ifnum #1\expandafter \@firstoftwo
 \else \expandafter \@secondoftwo
 \fi
}%
\providecommand \@ifx [1]{%
 \ifx #1\expandafter \@firstoftwo
 \else \expandafter \@secondoftwo
 \fi
}%
\providecommand \natexlab [1]{#1}%
\providecommand \enquote  [1]{``#1''}%
\providecommand \bibnamefont  [1]{#1}%
\providecommand \bibfnamefont [1]{#1}%
\providecommand \citenamefont [1]{#1}%
\providecommand \href@noop [0]{\@secondoftwo}%
\providecommand \href [0]{\begingroup \@sanitize@url \@href}%
\providecommand \@href[1]{\@@startlink{#1}\@@href}%
\providecommand \@@href[1]{\endgroup#1\@@endlink}%
\providecommand \@sanitize@url [0]{\catcode `\\12\catcode `\$12\catcode
  `\&12\catcode `\#12\catcode `\^12\catcode `\_12\catcode `\%12\relax}%
\providecommand \@@startlink[1]{}%
\providecommand \@@endlink[0]{}%
\providecommand \url  [0]{\begingroup\@sanitize@url \@url }%
\providecommand \@url [1]{\endgroup\@href {#1}{\urlprefix }}%
\providecommand \urlprefix  [0]{URL }%
\providecommand \Eprint [0]{\href }%
\providecommand \doibase [0]{http://dx.doi.org/}%
\providecommand \selectlanguage [0]{\@gobble}%
\providecommand \bibinfo  [0]{\@secondoftwo}%
\providecommand \bibfield  [0]{\@secondoftwo}%
\providecommand \translation [1]{[#1]}%
\providecommand \BibitemOpen [0]{}%
\providecommand \bibitemStop [0]{}%
\providecommand \bibitemNoStop [0]{.\EOS\space}%
\providecommand \EOS [0]{\spacefactor3000\relax}%
\providecommand \BibitemShut  [1]{\csname bibitem#1\endcsname}%
\let\auto@bib@innerbib\@empty
\bibitem [{\citenamefont {Gisin}\ \emph {et~al.}(2002)\citenamefont {Gisin},
  \citenamefont {Ribordy}, \citenamefont {Tittel},\ and\ \citenamefont
  {Zbinden}}]{gisin2002quantum}%
  \BibitemOpen
  \bibfield  {author} {\bibinfo {author} {\bibfnamefont {N.}~\bibnamefont
  {Gisin}}, \bibinfo {author} {\bibfnamefont {G.}~\bibnamefont {Ribordy}},
  \bibinfo {author} {\bibfnamefont {W.}~\bibnamefont {Tittel}}, \ and\ \bibinfo
  {author} {\bibfnamefont {H.}~\bibnamefont {Zbinden}},\ }\href@noop {}
  {\bibfield  {journal} {\bibinfo  {journal} {Rev. Mod. Phys.}\ }\textbf
  {\bibinfo {volume} {74}},\ \bibinfo {pages} {145} (\bibinfo {year}
  {2002})}\BibitemShut {NoStop}%
\bibitem [{\citenamefont {Pathak}(2013)}]{pathak2013elements}%
  \BibitemOpen
  \bibfield  {author} {\bibinfo {author} {\bibfnamefont {A.}~\bibnamefont
  {Pathak}},\ }\href@noop {} {\emph {\bibinfo {title} {Elements of quantum
  computation and quantum communication}}}\ (\bibinfo  {publisher} {Taylor \&
  Francis},\ \bibinfo {year} {2013})\BibitemShut {NoStop}%
\bibitem [{\citenamefont {Shenoy-Hejamadi}\ \emph {et~al.}(2017)\citenamefont
  {Shenoy-Hejamadi}, \citenamefont {Pathak},\ and\ \citenamefont
  {Radhakrishna}}]{shenoy2017quantum}%
  \BibitemOpen
  \bibfield  {author} {\bibinfo {author} {\bibfnamefont {A.}~\bibnamefont
  {Shenoy-Hejamadi}}, \bibinfo {author} {\bibfnamefont {A.}~\bibnamefont
  {Pathak}}, \ and\ \bibinfo {author} {\bibfnamefont {S.}~\bibnamefont
  {Radhakrishna}},\ }\href@noop {} {\bibfield  {journal} {\bibinfo  {journal}
  {Quanta}\ }\textbf {\bibinfo {volume} {6}},\ \bibinfo {pages} {1} (\bibinfo
  {year} {2017})}\BibitemShut {NoStop}%
\bibitem [{\citenamefont {Bennett}\ and\ \citenamefont
  {Brassard}(1984)}]{bennett1984quantum}%
  \BibitemOpen
  \bibfield  {author} {\bibinfo {author} {\bibfnamefont {C.~H.}\ \bibnamefont
  {Bennett}}\ and\ \bibinfo {author} {\bibfnamefont {G.}~\bibnamefont
  {Brassard}},\ }in\ \href@noop {} {\emph {\bibinfo {booktitle} {International
  Conference on Computer System and Signal Processing, IEEE, 1984}}}\ (\bibinfo
  {year} {1984})\ pp.\ \bibinfo {pages} {175--179}\BibitemShut {NoStop}%
\bibitem [{\citenamefont {Ekert}(1991)}]{ekert1991quantum}%
  \BibitemOpen
  \bibfield  {author} {\bibinfo {author} {\bibfnamefont {A.~K.}\ \bibnamefont
  {Ekert}},\ }\href@noop {} {\bibfield  {journal} {\bibinfo  {journal} {Phys.
  Rev. Lett.}\ }\textbf {\bibinfo {volume} {67}},\ \bibinfo {pages} {661}
  (\bibinfo {year} {1991})}\BibitemShut {NoStop}%
\bibitem [{\citenamefont {Bennett}(1992)}]{bennett1992quantum}%
  \BibitemOpen
  \bibfield  {author} {\bibinfo {author} {\bibfnamefont {C.~H.}\ \bibnamefont
  {Bennett}},\ }\href@noop {} {\bibfield  {journal} {\bibinfo  {journal} {Phys.
  Rev. Lett.}\ }\textbf {\bibinfo {volume} {68}},\ \bibinfo {pages} {3121}
  (\bibinfo {year} {1992})}\BibitemShut {NoStop}%
\bibitem [{\citenamefont {Chefles}(1998)}]{chefles1998unambiguous}%
  \BibitemOpen
  \bibfield  {author} {\bibinfo {author} {\bibfnamefont {A.}~\bibnamefont
  {Chefles}},\ }\href@noop {} {\bibfield  {journal} {\bibinfo  {journal} {Phys.
  Lett. A}\ }\textbf {\bibinfo {volume} {239}},\ \bibinfo {pages} {339}
  (\bibinfo {year} {1998})}\BibitemShut {NoStop}%
\bibitem [{\citenamefont {Du{\v{s}}ek}\ \emph {et~al.}(2006)\citenamefont
  {Du{\v{s}}ek}, \citenamefont {L{\"u}tkenhaus},\ and\ \citenamefont
  {Hendrych}}]{dusek2006quantum}%
  \BibitemOpen
  \bibfield  {author} {\bibinfo {author} {\bibfnamefont {M.}~\bibnamefont
  {Du{\v{s}}ek}}, \bibinfo {author} {\bibfnamefont {N.}~\bibnamefont
  {L{\"u}tkenhaus}}, \ and\ \bibinfo {author} {\bibfnamefont {M.}~\bibnamefont
  {Hendrych}},\ }\href@noop {} {\bibfield  {journal} {\bibinfo  {journal}
  {Progress in Optics}\ }\textbf {\bibinfo {volume} {49}},\ \bibinfo {pages}
  {381} (\bibinfo {year} {2006})}\BibitemShut {NoStop}%
\bibitem [{\citenamefont {Lucamarini}\ \emph {et~al.}(2009)\citenamefont
  {Lucamarini}, \citenamefont {Di~Giuseppe},\ and\ \citenamefont
  {Tamaki}}]{lucamarini2009robust}%
  \BibitemOpen
  \bibfield  {author} {\bibinfo {author} {\bibfnamefont {M.}~\bibnamefont
  {Lucamarini}}, \bibinfo {author} {\bibfnamefont {G.}~\bibnamefont
  {Di~Giuseppe}}, \ and\ \bibinfo {author} {\bibfnamefont {K.}~\bibnamefont
  {Tamaki}},\ }\href@noop {} {\bibfield  {journal} {\bibinfo  {journal} {Phys.
  Rev. A}\ }\textbf {\bibinfo {volume} {80}},\ \bibinfo {pages} {032327}
  (\bibinfo {year} {2009})}\BibitemShut {NoStop}%
\bibitem [{\citenamefont {Lucamarini}\ \emph {et~al.}(2012)\citenamefont
  {Lucamarini}, \citenamefont {Vallone}, \citenamefont {Gianani}, \citenamefont
  {Mataloni},\ and\ \citenamefont {Di~Giuseppe}}]{lucamarini2012device}%
  \BibitemOpen
  \bibfield  {author} {\bibinfo {author} {\bibfnamefont {M.}~\bibnamefont
  {Lucamarini}}, \bibinfo {author} {\bibfnamefont {G.}~\bibnamefont {Vallone}},
  \bibinfo {author} {\bibfnamefont {I.}~\bibnamefont {Gianani}}, \bibinfo
  {author} {\bibfnamefont {P.}~\bibnamefont {Mataloni}}, \ and\ \bibinfo
  {author} {\bibfnamefont {G.}~\bibnamefont {Di~Giuseppe}},\ }\href@noop {}
  {\bibfield  {journal} {\bibinfo  {journal} {Phys. Rev. A}\ }\textbf {\bibinfo
  {volume} {86}},\ \bibinfo {pages} {032325} (\bibinfo {year}
  {2012})}\BibitemShut {NoStop}%
\bibitem [{\citenamefont {Yang}\ \emph {et~al.}(2014)\citenamefont {Yang},
  \citenamefont {Sun}, \citenamefont {Xu},\ and\ \citenamefont
  {Tian}}]{yang2014flexible}%
  \BibitemOpen
  \bibfield  {author} {\bibinfo {author} {\bibfnamefont {Y.-G.}\ \bibnamefont
  {Yang}}, \bibinfo {author} {\bibfnamefont {S.-J.}\ \bibnamefont {Sun}},
  \bibinfo {author} {\bibfnamefont {P.}~\bibnamefont {Xu}}, \ and\ \bibinfo
  {author} {\bibfnamefont {J.}~\bibnamefont {Tian}},\ }\href@noop {} {\bibfield
   {journal} {\bibinfo  {journal} {Quantum Inf. Process.}\ }\textbf {\bibinfo
  {volume} {13}},\ \bibinfo {pages} {805} (\bibinfo {year} {2014})}\BibitemShut
  {NoStop}%
\bibitem [{\citenamefont {Chang}\ \emph {et~al.}(2017)\citenamefont {Chang},
  \citenamefont {Zhang},\ and\ \citenamefont {Zhu}}]{chang2017comment}%
  \BibitemOpen
  \bibfield  {author} {\bibinfo {author} {\bibfnamefont {Y.}~\bibnamefont
  {Chang}}, \bibinfo {author} {\bibfnamefont {S.-B.}\ \bibnamefont {Zhang}}, \
  and\ \bibinfo {author} {\bibfnamefont {J.-M.}\ \bibnamefont {Zhu}},\
  }\href@noop {} {\bibfield  {journal} {\bibinfo  {journal} {Quantum Inf.
  Process.}\ }\textbf {\bibinfo {volume} {16}},\ \bibinfo {pages} {86}
  (\bibinfo {year} {2017})}\BibitemShut {NoStop}%
\bibitem [{\citenamefont {Ko}\ \emph {et~al.}(2018)\citenamefont {Ko},
  \citenamefont {Choi}, \citenamefont {Choe},\ and\ \citenamefont
  {Youn}}]{ko2018advanced}%
  \BibitemOpen
  \bibfield  {author} {\bibinfo {author} {\bibfnamefont {H.}~\bibnamefont
  {Ko}}, \bibinfo {author} {\bibfnamefont {B.-S.}\ \bibnamefont {Choi}},
  \bibinfo {author} {\bibfnamefont {J.-S.}\ \bibnamefont {Choe}}, \ and\
  \bibinfo {author} {\bibfnamefont {C.~J.}\ \bibnamefont {Youn}},\ }\href@noop
  {} {\bibfield  {journal} {\bibinfo  {journal} {Quantum Inf. Process.}\
  }\textbf {\bibinfo {volume} {17}},\ \bibinfo {pages} {17} (\bibinfo {year}
  {2018})}\BibitemShut {NoStop}%
\bibitem [{\citenamefont {Braunstein}\ and\ \citenamefont
  {Van~Loock}(2005)}]{braunstein2005quantum}%
  \BibitemOpen
  \bibfield  {author} {\bibinfo {author} {\bibfnamefont {S.~L.}\ \bibnamefont
  {Braunstein}}\ and\ \bibinfo {author} {\bibfnamefont {P.}~\bibnamefont
  {Van~Loock}},\ }\href@noop {} {\bibfield  {journal} {\bibinfo  {journal}
  {Rev. Mod. Phys.}\ }\textbf {\bibinfo {volume} {77}},\ \bibinfo {pages} {513}
  (\bibinfo {year} {2005})}\BibitemShut {NoStop}%
\bibitem [{\citenamefont {Andersen}\ \emph {et~al.}(2010)\citenamefont
  {Andersen}, \citenamefont {Leuchs},\ and\ \citenamefont
  {Silberhorn}}]{andersen2010continuous}%
  \BibitemOpen
  \bibfield  {author} {\bibinfo {author} {\bibfnamefont {U.~L.}\ \bibnamefont
  {Andersen}}, \bibinfo {author} {\bibfnamefont {G.}~\bibnamefont {Leuchs}}, \
  and\ \bibinfo {author} {\bibfnamefont {C.}~\bibnamefont {Silberhorn}},\
  }\href@noop {} {\bibfield  {journal} {\bibinfo  {journal} {Laser \& Photonics
  Reviews}\ }\textbf {\bibinfo {volume} {4}},\ \bibinfo {pages} {337} (\bibinfo
  {year} {2010})}\BibitemShut {NoStop}%
\bibitem [{\citenamefont {Weedbrook}\ \emph {et~al.}(2012)\citenamefont
  {Weedbrook}, \citenamefont {Pirandola}, \citenamefont
  {Garc{\'\i}a-Patr{\'o}n}, \citenamefont {Cerf}, \citenamefont {Ralph},
  \citenamefont {Shapiro},\ and\ \citenamefont
  {Lloyd}}]{weedbrook2012gaussian}%
  \BibitemOpen
  \bibfield  {author} {\bibinfo {author} {\bibfnamefont {C.}~\bibnamefont
  {Weedbrook}}, \bibinfo {author} {\bibfnamefont {S.}~\bibnamefont
  {Pirandola}}, \bibinfo {author} {\bibfnamefont {R.}~\bibnamefont
  {Garc{\'\i}a-Patr{\'o}n}}, \bibinfo {author} {\bibfnamefont {N.~J.}\
  \bibnamefont {Cerf}}, \bibinfo {author} {\bibfnamefont {T.~C.}\ \bibnamefont
  {Ralph}}, \bibinfo {author} {\bibfnamefont {J.~H.}\ \bibnamefont {Shapiro}},
  \ and\ \bibinfo {author} {\bibfnamefont {S.}~\bibnamefont {Lloyd}},\
  }\href@noop {} {\bibfield  {journal} {\bibinfo  {journal} {Rev. Mod. Phys.}\
  }\textbf {\bibinfo {volume} {84}},\ \bibinfo {pages} {621} (\bibinfo {year}
  {2012})}\BibitemShut {NoStop}%
\bibitem [{\citenamefont {Hillery}(2000)}]{hillery2000quantum}%
  \BibitemOpen
  \bibfield  {author} {\bibinfo {author} {\bibfnamefont {M.}~\bibnamefont
  {Hillery}},\ }\href@noop {} {\bibfield  {journal} {\bibinfo  {journal} {Phys.
  Rev. A}\ }\textbf {\bibinfo {volume} {61}},\ \bibinfo {pages} {022309}
  (\bibinfo {year} {2000})}\BibitemShut {NoStop}%
\bibitem [{\citenamefont {Reid}(2000)}]{reid2000quantum}%
  \BibitemOpen
  \bibfield  {author} {\bibinfo {author} {\bibfnamefont {M.~D.}\ \bibnamefont
  {Reid}},\ }\href@noop {} {\bibfield  {journal} {\bibinfo  {journal} {Phys.
  Rev. A}\ }\textbf {\bibinfo {volume} {62}},\ \bibinfo {pages} {062308}
  (\bibinfo {year} {2000})}\BibitemShut {NoStop}%
\bibitem [{\citenamefont {Ralph}(1999)}]{ralph1999continuous}%
  \BibitemOpen
  \bibfield  {author} {\bibinfo {author} {\bibfnamefont {T.~C.}\ \bibnamefont
  {Ralph}},\ }\href@noop {} {\bibfield  {journal} {\bibinfo  {journal} {Phys.
  Rev. A}\ }\textbf {\bibinfo {volume} {61}},\ \bibinfo {pages} {010303}
  (\bibinfo {year} {1999})}\BibitemShut {NoStop}%
\bibitem [{\citenamefont {Ralph}(2000)}]{ralph2000security}%
  \BibitemOpen
  \bibfield  {author} {\bibinfo {author} {\bibfnamefont {T.~C.}\ \bibnamefont
  {Ralph}},\ }\href@noop {} {\bibfield  {journal} {\bibinfo  {journal} {Phys.
  Rev. A}\ }\textbf {\bibinfo {volume} {62}},\ \bibinfo {pages} {062306}
  (\bibinfo {year} {2000})}\BibitemShut {NoStop}%
\bibitem [{\citenamefont {Gottesman}\ and\ \citenamefont
  {Preskill}(2001)}]{PhysRevA.63.022309}%
  \BibitemOpen
  \bibfield  {author} {\bibinfo {author} {\bibfnamefont {D.}~\bibnamefont
  {Gottesman}}\ and\ \bibinfo {author} {\bibfnamefont {J.}~\bibnamefont
  {Preskill}},\ }\href {\doibase 10.1103/PhysRevA.63.022309} {\bibfield
  {journal} {\bibinfo  {journal} {Phys. Rev. A}\ }\textbf {\bibinfo {volume}
  {63}},\ \bibinfo {pages} {022309} (\bibinfo {year} {2001})}\BibitemShut
  {NoStop}%
\bibitem [{\citenamefont {Grosshans}\ and\ \citenamefont
  {Grangier}(2002)}]{grosshans2002continuous}%
  \BibitemOpen
  \bibfield  {author} {\bibinfo {author} {\bibfnamefont {F.}~\bibnamefont
  {Grosshans}}\ and\ \bibinfo {author} {\bibfnamefont {P.}~\bibnamefont
  {Grangier}},\ }\href@noop {} {\bibfield  {journal} {\bibinfo  {journal}
  {Phys. Rev. Lett.}\ }\textbf {\bibinfo {volume} {88}},\ \bibinfo {pages}
  {057902} (\bibinfo {year} {2002})}\BibitemShut {NoStop}%
\bibitem [{\citenamefont {Cerf}\ \emph {et~al.}(2001)\citenamefont {Cerf},
  \citenamefont {Levy},\ and\ \citenamefont {Van~Assche}}]{cerf2001quantum}%
  \BibitemOpen
  \bibfield  {author} {\bibinfo {author} {\bibfnamefont {N.~J.}\ \bibnamefont
  {Cerf}}, \bibinfo {author} {\bibfnamefont {M.}~\bibnamefont {Levy}}, \ and\
  \bibinfo {author} {\bibfnamefont {G.}~\bibnamefont {Van~Assche}},\
  }\href@noop {} {\bibfield  {journal} {\bibinfo  {journal} {Phys. Rev. A}\
  }\textbf {\bibinfo {volume} {63}},\ \bibinfo {pages} {052311} (\bibinfo
  {year} {2001})}\BibitemShut {NoStop}%
\bibitem [{\citenamefont {L{\"u}tkenhaus}\ and\ \citenamefont
  {Shields}(2009)}]{lutkenhaus2009focus}%
  \BibitemOpen
  \bibfield  {author} {\bibinfo {author} {\bibfnamefont {N.}~\bibnamefont
  {L{\"u}tkenhaus}}\ and\ \bibinfo {author} {\bibfnamefont {A.}~\bibnamefont
  {Shields}},\ }\href@noop {} {\bibfield  {journal} {\bibinfo  {journal} {New
  J. Phys.}\ }\textbf {\bibinfo {volume} {11}},\ \bibinfo {pages} {045005}
  (\bibinfo {year} {2009})}\BibitemShut {NoStop}%
\bibitem [{\citenamefont {Diamanti}\ and\ \citenamefont
  {Leverrier}(2015)}]{diamanti2015distributing}%
  \BibitemOpen
  \bibfield  {author} {\bibinfo {author} {\bibfnamefont {E.}~\bibnamefont
  {Diamanti}}\ and\ \bibinfo {author} {\bibfnamefont {A.}~\bibnamefont
  {Leverrier}},\ }\href@noop {} {\bibfield  {journal} {\bibinfo  {journal}
  {Entropy}\ }\textbf {\bibinfo {volume} {17}},\ \bibinfo {pages} {6072}
  (\bibinfo {year} {2015})}\BibitemShut {NoStop}%
\bibitem [{\citenamefont {Borelli}\ \emph {et~al.}(2016)\citenamefont
  {Borelli}, \citenamefont {Aguiar}, \citenamefont {Roversi},\ and\
  \citenamefont {Vidiella-Barranco}}]{borelli2016quantum}%
  \BibitemOpen
  \bibfield  {author} {\bibinfo {author} {\bibfnamefont {L.~F.}\ \bibnamefont
  {Borelli}}, \bibinfo {author} {\bibfnamefont {L.~d.~S.}\ \bibnamefont
  {Aguiar}}, \bibinfo {author} {\bibfnamefont {J.~A.}\ \bibnamefont {Roversi}},
  \ and\ \bibinfo {author} {\bibfnamefont {A.}~\bibnamefont
  {Vidiella-Barranco}},\ }\href@noop {} {\bibfield  {journal} {\bibinfo
  {journal} {Quantum Inf. Process.}\ }\textbf {\bibinfo {volume} {15}},\
  \bibinfo {pages} {893} (\bibinfo {year} {2016})}\BibitemShut {NoStop}%
\bibitem [{\citenamefont {Scarani}\ \emph {et~al.}(2009)\citenamefont
  {Scarani}, \citenamefont {Bechmann-Pasquinucci}, \citenamefont {Cerf},
  \citenamefont {Du{\v{s}}ek}, \citenamefont {L{\"u}tkenhaus},\ and\
  \citenamefont {Peev}}]{scarani2009security}%
  \BibitemOpen
  \bibfield  {author} {\bibinfo {author} {\bibfnamefont {V.}~\bibnamefont
  {Scarani}}, \bibinfo {author} {\bibfnamefont {H.}~\bibnamefont
  {Bechmann-Pasquinucci}}, \bibinfo {author} {\bibfnamefont {N.~J.}\
  \bibnamefont {Cerf}}, \bibinfo {author} {\bibfnamefont {M.}~\bibnamefont
  {Du{\v{s}}ek}}, \bibinfo {author} {\bibfnamefont {N.}~\bibnamefont
  {L{\"u}tkenhaus}}, \ and\ \bibinfo {author} {\bibfnamefont {M.}~\bibnamefont
  {Peev}},\ }\href@noop {} {\bibfield  {journal} {\bibinfo  {journal} {Rev.
  Mod. Phys.}\ }\textbf {\bibinfo {volume} {81}},\ \bibinfo {pages} {1301}
  (\bibinfo {year} {2009})}\BibitemShut {NoStop}%
\bibitem [{\citenamefont {Leverrier}(2017)}]{leverrier2017security}%
  \BibitemOpen
  \bibfield  {author} {\bibinfo {author} {\bibfnamefont {A.}~\bibnamefont
  {Leverrier}},\ }\href@noop {} {\bibfield  {journal} {\bibinfo  {journal}
  {Phys. Rev. Lett.}\ }\textbf {\bibinfo {volume} {118}},\ \bibinfo {pages}
  {200501} (\bibinfo {year} {2017})}\BibitemShut {NoStop}%
\bibitem [{\citenamefont {Leverrier}(2015)}]{leverrier2015composable}%
  \BibitemOpen
  \bibfield  {author} {\bibinfo {author} {\bibfnamefont {A.}~\bibnamefont
  {Leverrier}},\ }\href@noop {} {\bibfield  {journal} {\bibinfo  {journal}
  {Phys. Rev. Lett.}\ }\textbf {\bibinfo {volume} {114}},\ \bibinfo {pages}
  {070501} (\bibinfo {year} {2015})}\BibitemShut {NoStop}%
\bibitem [{\citenamefont {Liu}\ \emph {et~al.}(2018)\citenamefont {Liu},
  \citenamefont {Huang}, \citenamefont {Peng}, \citenamefont {Fan},\ and\
  \citenamefont {Zeng}}]{liu2018integrating}%
  \BibitemOpen
  \bibfield  {author} {\bibinfo {author} {\bibfnamefont {W.}~\bibnamefont
  {Liu}}, \bibinfo {author} {\bibfnamefont {P.}~\bibnamefont {Huang}}, \bibinfo
  {author} {\bibfnamefont {J.}~\bibnamefont {Peng}}, \bibinfo {author}
  {\bibfnamefont {J.}~\bibnamefont {Fan}}, \ and\ \bibinfo {author}
  {\bibfnamefont {G.}~\bibnamefont {Zeng}},\ }\href@noop {} {\bibfield
  {journal} {\bibinfo  {journal} {Phys. Rev. A}\ }\textbf {\bibinfo {volume}
  {97}},\ \bibinfo {pages} {022316} (\bibinfo {year} {2018})}\BibitemShut
  {NoStop}%
\bibitem [{\citenamefont {Lupo}\ \emph {et~al.}(2018)\citenamefont {Lupo},
  \citenamefont {Ottaviani}, \citenamefont {Papanastasiou},\ and\ \citenamefont
  {Pirandola}}]{lupo2018parameter}%
  \BibitemOpen
  \bibfield  {author} {\bibinfo {author} {\bibfnamefont {C.}~\bibnamefont
  {Lupo}}, \bibinfo {author} {\bibfnamefont {C.}~\bibnamefont {Ottaviani}},
  \bibinfo {author} {\bibfnamefont {P.}~\bibnamefont {Papanastasiou}}, \ and\
  \bibinfo {author} {\bibfnamefont {S.}~\bibnamefont {Pirandola}},\ }\href@noop
  {} {\bibfield  {journal} {\bibinfo  {journal} {Phys. Rev. Lett.}\ }\textbf
  {\bibinfo {volume} {120}},\ \bibinfo {pages} {220505} (\bibinfo {year}
  {2018})}\BibitemShut {NoStop}%
\bibitem [{\citenamefont {Pirandola}\ \emph {et~al.}(2015)\citenamefont
  {Pirandola}, \citenamefont {Ottaviani}, \citenamefont {Spedalieri},
  \citenamefont {Weedbrook}, \citenamefont {Braunstein}, \citenamefont {Lloyd},
  \citenamefont {Gehring}, \citenamefont {Jacobsen},\ and\ \citenamefont
  {Andersen}}]{pirandola2015high}%
  \BibitemOpen
  \bibfield  {author} {\bibinfo {author} {\bibfnamefont {S.}~\bibnamefont
  {Pirandola}}, \bibinfo {author} {\bibfnamefont {C.}~\bibnamefont
  {Ottaviani}}, \bibinfo {author} {\bibfnamefont {G.}~\bibnamefont
  {Spedalieri}}, \bibinfo {author} {\bibfnamefont {C.}~\bibnamefont
  {Weedbrook}}, \bibinfo {author} {\bibfnamefont {S.~L.}\ \bibnamefont
  {Braunstein}}, \bibinfo {author} {\bibfnamefont {S.}~\bibnamefont {Lloyd}},
  \bibinfo {author} {\bibfnamefont {T.}~\bibnamefont {Gehring}}, \bibinfo
  {author} {\bibfnamefont {C.~S.}\ \bibnamefont {Jacobsen}}, \ and\ \bibinfo
  {author} {\bibfnamefont {U.~L.}\ \bibnamefont {Andersen}},\ }\href@noop {}
  {\bibfield  {journal} {\bibinfo  {journal} {Nature Photonics}\ }\textbf
  {\bibinfo {volume} {9}},\ \bibinfo {pages} {397} (\bibinfo {year}
  {2015})}\BibitemShut {NoStop}%
\bibitem [{\citenamefont {Marshall}\ and\ \citenamefont
  {Weedbrook}(2014)}]{marshall2014device}%
  \BibitemOpen
  \bibfield  {author} {\bibinfo {author} {\bibfnamefont {K.}~\bibnamefont
  {Marshall}}\ and\ \bibinfo {author} {\bibfnamefont {C.}~\bibnamefont
  {Weedbrook}},\ }\href@noop {} {\bibfield  {journal} {\bibinfo  {journal}
  {Phys. Rev. A}\ }\textbf {\bibinfo {volume} {90}},\ \bibinfo {pages} {042311}
  (\bibinfo {year} {2014})}\BibitemShut {NoStop}%
\bibitem [{\citenamefont {Zhou}\ \emph {et~al.}(2018)\citenamefont {Zhou},
  \citenamefont {Huang},\ and\ \citenamefont {Guo}}]{zhou2018long}%
  \BibitemOpen
  \bibfield  {author} {\bibinfo {author} {\bibfnamefont {J.}~\bibnamefont
  {Zhou}}, \bibinfo {author} {\bibfnamefont {D.}~\bibnamefont {Huang}}, \ and\
  \bibinfo {author} {\bibfnamefont {Y.}~\bibnamefont {Guo}},\ }\href@noop {}
  {\bibfield  {journal} {\bibinfo  {journal} {Phys. Rev. A}\ }\textbf {\bibinfo
  {volume} {98}},\ \bibinfo {pages} {042303} (\bibinfo {year}
  {2018})}\BibitemShut {NoStop}%
\bibitem [{\citenamefont {Heim}\ \emph {et~al.}(2014)\citenamefont {Heim},
  \citenamefont {Peuntinger}, \citenamefont {Killoran}, \citenamefont {Khan},
  \citenamefont {Wittmann}, \citenamefont {Marquardt},\ and\ \citenamefont
  {Leuchs}}]{heim2014atmospheric}%
  \BibitemOpen
  \bibfield  {author} {\bibinfo {author} {\bibfnamefont {B.}~\bibnamefont
  {Heim}}, \bibinfo {author} {\bibfnamefont {C.}~\bibnamefont {Peuntinger}},
  \bibinfo {author} {\bibfnamefont {N.}~\bibnamefont {Killoran}}, \bibinfo
  {author} {\bibfnamefont {I.}~\bibnamefont {Khan}}, \bibinfo {author}
  {\bibfnamefont {C.}~\bibnamefont {Wittmann}}, \bibinfo {author}
  {\bibfnamefont {C.}~\bibnamefont {Marquardt}}, \ and\ \bibinfo {author}
  {\bibfnamefont {G.}~\bibnamefont {Leuchs}},\ }\href@noop {} {\bibfield
  {journal} {\bibinfo  {journal} {New J. Phys.}\ }\textbf {\bibinfo {volume}
  {16}},\ \bibinfo {pages} {113018} (\bibinfo {year} {2014})}\BibitemShut
  {NoStop}%
\bibitem [{\citenamefont {Hosseinidehaj}\ \emph {et~al.}(2018)\citenamefont
  {Hosseinidehaj}, \citenamefont {Babar}, \citenamefont {Malaney},
  \citenamefont {Ng},\ and\ \citenamefont
  {Hanzo}}]{hosseinidehaj2018satellite}%
  \BibitemOpen
  \bibfield  {author} {\bibinfo {author} {\bibfnamefont {N.}~\bibnamefont
  {Hosseinidehaj}}, \bibinfo {author} {\bibfnamefont {Z.}~\bibnamefont
  {Babar}}, \bibinfo {author} {\bibfnamefont {R.}~\bibnamefont {Malaney}},
  \bibinfo {author} {\bibfnamefont {S.~X.}\ \bibnamefont {Ng}}, \ and\ \bibinfo
  {author} {\bibfnamefont {L.}~\bibnamefont {Hanzo}},\ }\href@noop {}
  {\bibfield  {journal} {\bibinfo  {journal} {IEEE Communications Surveys \&
  Tutorials}\ }\textbf {\bibinfo {volume} {21}},\ \bibinfo {pages} {881}
  (\bibinfo {year} {2018})}\BibitemShut {NoStop}%
\bibitem [{\citenamefont {Liu}\ \emph {et~al.}(2015)\citenamefont {Liu},
  \citenamefont {Peng}, \citenamefont {Wang}, \citenamefont {Cao},
  \citenamefont {Huang}, \citenamefont {Lin}, \citenamefont {Huang},\ and\
  \citenamefont {Zeng}}]{liu2015hybrid}%
  \BibitemOpen
  \bibfield  {author} {\bibinfo {author} {\bibfnamefont {W.}~\bibnamefont
  {Liu}}, \bibinfo {author} {\bibfnamefont {J.}~\bibnamefont {Peng}}, \bibinfo
  {author} {\bibfnamefont {C.}~\bibnamefont {Wang}}, \bibinfo {author}
  {\bibfnamefont {Z.}~\bibnamefont {Cao}}, \bibinfo {author} {\bibfnamefont
  {D.}~\bibnamefont {Huang}}, \bibinfo {author} {\bibfnamefont
  {D.}~\bibnamefont {Lin}}, \bibinfo {author} {\bibfnamefont {P.}~\bibnamefont
  {Huang}}, \ and\ \bibinfo {author} {\bibfnamefont {G.}~\bibnamefont {Zeng}},\
  }\href@noop {} {\bibfield  {journal} {\bibinfo  {journal} {SCIENCE CHINA
  Physics, Mechanics \& Astronomy}\ }\textbf {\bibinfo {volume} {58}},\
  \bibinfo {pages} {1} (\bibinfo {year} {2015})}\BibitemShut {NoStop}%
\bibitem [{\citenamefont {Jacobsen}\ \emph {et~al.}(2018)\citenamefont
  {Jacobsen}, \citenamefont {Madsen}, \citenamefont {Usenko}, \citenamefont
  {Filip},\ and\ \citenamefont {Andersen}}]{jacobsen2018complete}%
  \BibitemOpen
  \bibfield  {author} {\bibinfo {author} {\bibfnamefont {C.~S.}\ \bibnamefont
  {Jacobsen}}, \bibinfo {author} {\bibfnamefont {L.~S.}\ \bibnamefont
  {Madsen}}, \bibinfo {author} {\bibfnamefont {V.~C.}\ \bibnamefont {Usenko}},
  \bibinfo {author} {\bibfnamefont {R.}~\bibnamefont {Filip}}, \ and\ \bibinfo
  {author} {\bibfnamefont {U.~L.}\ \bibnamefont {Andersen}},\ }\href@noop {}
  {\bibfield  {journal} {\bibinfo  {journal} {npj Quantum Inf.}\ }\textbf
  {\bibinfo {volume} {4}},\ \bibinfo {pages} {32} (\bibinfo {year}
  {2018})}\BibitemShut {NoStop}%
\bibitem [{\citenamefont {Croal}\ \emph {et~al.}(2016)\citenamefont {Croal},
  \citenamefont {Peuntinger}, \citenamefont {Heim}, \citenamefont {Khan},
  \citenamefont {Marquardt}, \citenamefont {Leuchs}, \citenamefont {Wallden},
  \citenamefont {Andersson},\ and\ \citenamefont {Korolkova}}]{croal2016free}%
  \BibitemOpen
  \bibfield  {author} {\bibinfo {author} {\bibfnamefont {C.}~\bibnamefont
  {Croal}}, \bibinfo {author} {\bibfnamefont {C.}~\bibnamefont {Peuntinger}},
  \bibinfo {author} {\bibfnamefont {B.}~\bibnamefont {Heim}}, \bibinfo {author}
  {\bibfnamefont {I.}~\bibnamefont {Khan}}, \bibinfo {author} {\bibfnamefont
  {C.}~\bibnamefont {Marquardt}}, \bibinfo {author} {\bibfnamefont
  {G.}~\bibnamefont {Leuchs}}, \bibinfo {author} {\bibfnamefont
  {P.}~\bibnamefont {Wallden}}, \bibinfo {author} {\bibfnamefont
  {E.}~\bibnamefont {Andersson}}, \ and\ \bibinfo {author} {\bibfnamefont
  {N.}~\bibnamefont {Korolkova}},\ }\href@noop {} {\bibfield  {journal}
  {\bibinfo  {journal} {Phys. Rev. Lett.}\ }\textbf {\bibinfo {volume} {117}},\
  \bibinfo {pages} {100503} (\bibinfo {year} {2016})}\BibitemShut {NoStop}%
\bibitem [{\citenamefont {Saxena}\ \emph {et~al.}(2019)\citenamefont {Saxena},
  \citenamefont {Thapliyal},\ and\ \citenamefont
  {Pathak}}]{saxena2019continuous}%
  \BibitemOpen
  \bibfield  {author} {\bibinfo {author} {\bibfnamefont {A.}~\bibnamefont
  {Saxena}}, \bibinfo {author} {\bibfnamefont {K.}~\bibnamefont {Thapliyal}}, \
  and\ \bibinfo {author} {\bibfnamefont {A.}~\bibnamefont {Pathak}},\
  }\href@noop {} {\bibfield  {journal} {\bibinfo  {journal} {arXiv preprint
  arXiv:1902.00458}\ } (\bibinfo {year} {2019})}\BibitemShut {NoStop}%
\bibitem [{\citenamefont {Wu}\ \emph {et~al.}(2016)\citenamefont {Wu},
  \citenamefont {Zhou}, \citenamefont {Gong}, \citenamefont {Guo},
  \citenamefont {Zhang},\ and\ \citenamefont {He}}]{wu2016continuous}%
  \BibitemOpen
  \bibfield  {author} {\bibinfo {author} {\bibfnamefont {Y.}~\bibnamefont
  {Wu}}, \bibinfo {author} {\bibfnamefont {J.}~\bibnamefont {Zhou}}, \bibinfo
  {author} {\bibfnamefont {X.}~\bibnamefont {Gong}}, \bibinfo {author}
  {\bibfnamefont {Y.}~\bibnamefont {Guo}}, \bibinfo {author} {\bibfnamefont
  {Z.-M.}\ \bibnamefont {Zhang}}, \ and\ \bibinfo {author} {\bibfnamefont
  {G.}~\bibnamefont {He}},\ }\href@noop {} {\bibfield  {journal} {\bibinfo
  {journal} {Phys. Rev. A}\ }\textbf {\bibinfo {volume} {93}},\ \bibinfo
  {pages} {022325} (\bibinfo {year} {2016})}\BibitemShut {NoStop}%
\bibitem [{\citenamefont {Qi}\ and\ \citenamefont
  {Siopsis}(2015)}]{qi2015loss}%
  \BibitemOpen
  \bibfield  {author} {\bibinfo {author} {\bibfnamefont {B.}~\bibnamefont
  {Qi}}\ and\ \bibinfo {author} {\bibfnamefont {G.}~\bibnamefont {Siopsis}},\
  }\href@noop {} {\bibfield  {journal} {\bibinfo  {journal} {Phys. Rev. A}\
  }\textbf {\bibinfo {volume} {91}},\ \bibinfo {pages} {042337} (\bibinfo
  {year} {2015})}\BibitemShut {NoStop}%
\bibitem [{\citenamefont {Ma}\ \emph {et~al.}(2019)\citenamefont {Ma},
  \citenamefont {Huang}, \citenamefont {Bai}, \citenamefont {Wang},
  \citenamefont {Wang}, \citenamefont {Bao},\ and\ \citenamefont
  {Zeng}}]{ma2019long}%
  \BibitemOpen
  \bibfield  {author} {\bibinfo {author} {\bibfnamefont {H.-X.}\ \bibnamefont
  {Ma}}, \bibinfo {author} {\bibfnamefont {P.}~\bibnamefont {Huang}}, \bibinfo
  {author} {\bibfnamefont {D.-Y.}\ \bibnamefont {Bai}}, \bibinfo {author}
  {\bibfnamefont {T.}~\bibnamefont {Wang}}, \bibinfo {author} {\bibfnamefont
  {S.-Y.}\ \bibnamefont {Wang}}, \bibinfo {author} {\bibfnamefont {W.-S.}\
  \bibnamefont {Bao}}, \ and\ \bibinfo {author} {\bibfnamefont {G.-H.}\
  \bibnamefont {Zeng}},\ }\href@noop {} {\bibfield  {journal} {\bibinfo
  {journal} {Phys. Rev. A}\ }\textbf {\bibinfo {volume} {99}},\ \bibinfo
  {pages} {022322} (\bibinfo {year} {2019})}\BibitemShut {NoStop}%
\bibitem [{\citenamefont {Parigi}\ \emph {et~al.}(2007)\citenamefont {Parigi},
  \citenamefont {Zavatta}, \citenamefont {Kim},\ and\ \citenamefont
  {Bellini}}]{parigi2007probing}%
  \BibitemOpen
  \bibfield  {author} {\bibinfo {author} {\bibfnamefont {V.}~\bibnamefont
  {Parigi}}, \bibinfo {author} {\bibfnamefont {A.}~\bibnamefont {Zavatta}},
  \bibinfo {author} {\bibfnamefont {M.}~\bibnamefont {Kim}}, \ and\ \bibinfo
  {author} {\bibfnamefont {M.}~\bibnamefont {Bellini}},\ }\href@noop {}
  {\bibfield  {journal} {\bibinfo  {journal} {Science}\ }\textbf {\bibinfo
  {volume} {317}},\ \bibinfo {pages} {1890} (\bibinfo {year}
  {2007})}\BibitemShut {NoStop}%
\bibitem [{\citenamefont {Thapliyal}\ \emph {et~al.}(2017)\citenamefont
  {Thapliyal}, \citenamefont {Samantray}, \citenamefont {Banerji},\ and\
  \citenamefont {Pathak}}]{thapliyal2017comparison}%
  \BibitemOpen
  \bibfield  {author} {\bibinfo {author} {\bibfnamefont {K.}~\bibnamefont
  {Thapliyal}}, \bibinfo {author} {\bibfnamefont {N.~L.}\ \bibnamefont
  {Samantray}}, \bibinfo {author} {\bibfnamefont {J.}~\bibnamefont {Banerji}},
  \ and\ \bibinfo {author} {\bibfnamefont {A.}~\bibnamefont {Pathak}},\
  }\href@noop {} {\bibfield  {journal} {\bibinfo  {journal} {Phys. Lett. A}\
  }\textbf {\bibinfo {volume} {381}},\ \bibinfo {pages} {3178 } (\bibinfo
  {year} {2017})}\BibitemShut {NoStop}%
\bibitem [{\citenamefont {Malpani}\ \emph {et~al.}(2019)\citenamefont
  {Malpani}, \citenamefont {Alam}, \citenamefont {Thapliyal}, \citenamefont
  {Pathak}, \citenamefont {Narayanan},\ and\ \citenamefont
  {Banerjee}}]{malpani2018lower}%
  \BibitemOpen
  \bibfield  {author} {\bibinfo {author} {\bibfnamefont {P.}~\bibnamefont
  {Malpani}}, \bibinfo {author} {\bibfnamefont {N.}~\bibnamefont {Alam}},
  \bibinfo {author} {\bibfnamefont {K.}~\bibnamefont {Thapliyal}}, \bibinfo
  {author} {\bibfnamefont {A.}~\bibnamefont {Pathak}}, \bibinfo {author}
  {\bibfnamefont {V.}~\bibnamefont {Narayanan}}, \ and\ \bibinfo {author}
  {\bibfnamefont {S.}~\bibnamefont {Banerjee}},\ }\href@noop {} {\bibfield
  {journal} {\bibinfo  {journal} {Ann. Phys. (Berl.)}\ }\textbf {\bibinfo
  {volume} {531}},\ \bibinfo {pages} {1800318} (\bibinfo {year}
  {2019})}\BibitemShut {NoStop}%
\bibitem [{\citenamefont {Wigner}(1932)}]{wigner1932quantum}%
  \BibitemOpen
  \bibfield  {author} {\bibinfo {author} {\bibfnamefont {E.~P.}\ \bibnamefont
  {Wigner}},\ }\href@noop {} {\bibfield  {journal} {\bibinfo  {journal} {Phys.
  Rev.}\ }\textbf {\bibinfo {volume} {40}},\ \bibinfo {pages} {749} (\bibinfo
  {year} {1932})}\BibitemShut {NoStop}%
\bibitem [{\citenamefont {Glauber}(1963)}]{glauber1963coherent}%
  \BibitemOpen
  \bibfield  {author} {\bibinfo {author} {\bibfnamefont {R.~J.}\ \bibnamefont
  {Glauber}},\ }\href@noop {} {\bibfield  {journal} {\bibinfo  {journal} {Phys.
  Rev.}\ }\textbf {\bibinfo {volume} {131}},\ \bibinfo {pages} {2766} (\bibinfo
  {year} {1963})}\BibitemShut {NoStop}%
\bibitem [{\citenamefont {Sudarshan}(1963)}]{sudarshan1963equivalence}%
  \BibitemOpen
  \bibfield  {author} {\bibinfo {author} {\bibfnamefont {E.~C.~G.}\
  \bibnamefont {Sudarshan}},\ }\href@noop {} {\bibfield  {journal} {\bibinfo
  {journal} {Phys. Rev. Lett.}\ }\textbf {\bibinfo {volume} {10}},\ \bibinfo
  {pages} {277} (\bibinfo {year} {1963})}\BibitemShut {NoStop}%
\bibitem [{\citenamefont {Moya-Cessa}\ and\ \citenamefont
  {Knight}(1993)}]{moya1993series}%
  \BibitemOpen
  \bibfield  {author} {\bibinfo {author} {\bibfnamefont {H.}~\bibnamefont
  {Moya-Cessa}}\ and\ \bibinfo {author} {\bibfnamefont {P.~L.}\ \bibnamefont
  {Knight}},\ }\href@noop {} {\bibfield  {journal} {\bibinfo  {journal} {Phys.
  Rev. A}\ }\textbf {\bibinfo {volume} {48}},\ \bibinfo {pages} {2479}
  (\bibinfo {year} {1993})}\BibitemShut {NoStop}%
\bibitem [{\citenamefont {Horak}(2004)}]{horak2004role}%
  \BibitemOpen
  \bibfield  {author} {\bibinfo {author} {\bibfnamefont {P.}~\bibnamefont
  {Horak}},\ }\href@noop {} {\bibfield  {journal} {\bibinfo  {journal} {J. Mod.
  Opt.}\ }\textbf {\bibinfo {volume} {51}},\ \bibinfo {pages} {1249} (\bibinfo
  {year} {2004})}\BibitemShut {NoStop}%
\bibitem [{\citenamefont {L{\"u}tkenhaus}(1996)}]{lutkenhaus1996security}%
  \BibitemOpen
  \bibfield  {author} {\bibinfo {author} {\bibfnamefont {N.}~\bibnamefont
  {L{\"u}tkenhaus}},\ }\href@noop {} {\bibfield  {journal} {\bibinfo  {journal}
  {Phys. Rev. A}\ }\textbf {\bibinfo {volume} {54}},\ \bibinfo {pages} {97}
  (\bibinfo {year} {1996})}\BibitemShut {NoStop}%
\bibitem [{\citenamefont {Namiki}\ and\ \citenamefont
  {Hirano}(2004)}]{namiki2004practical}%
  \BibitemOpen
  \bibfield  {author} {\bibinfo {author} {\bibfnamefont {R.}~\bibnamefont
  {Namiki}}\ and\ \bibinfo {author} {\bibfnamefont {T.}~\bibnamefont
  {Hirano}},\ }\href@noop {} {\bibfield  {journal} {\bibinfo  {journal} {Phys.
  Rev. Lett.}\ }\textbf {\bibinfo {volume} {92}},\ \bibinfo {pages} {117901}
  (\bibinfo {year} {2004})}\BibitemShut {NoStop}%
\end{thebibliography}%

\end{document}